\documentclass[12pt]{amsart}
\usepackage{amscd}
\newtheoremstyle{xplain}
   {5ex}                 
   {5ex}                 
   {\itshape}            
   {}                    
   {\bf}                 
   {.}                   
   {0.5em}               
   {}                    
\theoremstyle{xplain}
\newtheorem{theorem}{Theorem}[section]
\newtheorem{lemma}[theorem]{Lemma}
\newtheorem{sublemma}[theorem]{Sublemma}

\newtheoremstyle{xdefinition}
   {5ex}                 
   {5ex}                 
   {}                    
   {}                    
   {\bf}                 
   {.}                   
   {0.5em}               
   {}                    
\theoremstyle{xdefinition}

\theoremstyle{remark}

\newsavebox{\SmallMathBox}

\def\operatorname#1{{\rm#1\,}}
\def\Trace{\operatorname{Tr}}
\def\text#1{{\hbox{#1}}}

\def\qedbox{\hbox{$\rlap{$\sqcap$}\sqcup$}}

\def\beq{\begin{eqnarray}}
\def\eeq{\end{eqnarray}}
\def\bea{\begin{array}}
\def\eea{\end{array}}
\def\nmonth{\ifcase\month\ \or January\or
February\or March\or April\or May\or June\or July\or August\or 
September\or October\or November\else December\fi}
\newcommand{\nn}{\nonumber}
\newcommand{\nats}{\mbox{${\rm I\!N }$}}
\def\pbgT{{\mathbb T}}

\begin{document}
\setcounter{page}{1}

\title{Heat asymptotics with spectral boundary conditions}

\author{Stuart Dowker,
     Peter Gilkey, and Klaus Kirsten}

\address{SD: Room 6.23 Dept Physics and Astronomy, The University of\newline
     Manchester, Oxford Road, Manchester UK M13 9PL\newline
    email:{\tt dowker@a13.ph.man.ac.uk};\newline
    {\tt http://theory.ph.man.ac.uk/theory/stuart.html}
    \newline\indent
    PG: Mathematics Department, University of Oregon,\newline Eugene Or 97403
    USA email:{\tt gilkey@darkwing.uoregon.edu};\newline
     {\tt http://darkwing.uoregon.edu/$\sim$gilkey}\newline\indent
    KK: Universit\"at Leipzig, Institut f\"ur Theoretische Physik,\newline
     Augustusplatz 10, 04109 Leipzig Deutschland\newline
     {\tt email: kirsten@itp.uni-leipzig.de}}


\thanks{SD: Research Partially supported by EPSRC Grant GR/L75708}
\thanks{PG: Research Partially supported by the NSF (USA), ESI
    (Austria), and MPI (Germany)}
\thanks{KK: Research Partially supported by DFG Grant B01112/4-2}
\thanks{subject class: Primary 58G25; PACS
             numbers: 1100, 0230, 0462}
\thanks{This paper is in final form and no version of it will be
submitted for publication elsewhere.}

\begin{abstract} Let $P$ be an operator of Dirac type on a compact
   Riemannian manifold with smooth boundary. We impose
   spectral boundary conditions and study the asymptotics of the
   heat trace of the associated operator of Laplace type.
\end{abstract}

\maketitle
\section{Introduction}
Let $M$ be a compact Riemannian manifold of dimension $m\ge3$ with smooth
boundary $\partial M$. Let $E_i$ be unitary bundles over $M$ and let
\begin{equation} P:C^\infty(E_1)\rightarrow C^\infty(E_2)\label{EQNu}\end{equation}
where $P$ is first order partial differential operator. Let $P^*$ be the formal adjoint
of $P$. We say display (\ref{EQNu}) is an elliptic complex of
\it Dirac type
\rm if the associated second order operators 
$$D_1:=P^*P\text{ and }D_2:=PP^*$$
on $C^\infty(E_1)$ and on $C^\infty(E_2)$ are of \it Laplace type \rm - i.e. if these
operators have scalar leading symbol given by the metric tensor. If $E_1=E_2$ and if
$P=P^*$, then
$P$ is said to be an operator of Dirac type; it is convenient, however, to work in this
slightly more general context.

We impose \it
spectral boundary conditions\rm; these were first introduced by Atiyah,
Patodi, and Singer \cite{APS} in their study of the index theorem for
manifolds with boundary. Let $\gamma$ be the leading symbol of the operator
$P$. Then
$\gamma+\gamma^*$ defines a unitary Clifford module structure on $E_1\oplus E_2$.
We choose a unitary connection $\nabla=\nabla_1\oplus\nabla_2$ on
$E_1\oplus E_2$ so that 
\begin{equation}\nabla(\gamma+\gamma^*)=0 \text{ and }
  (\nabla s,\tilde s)+(s,\nabla\tilde s)=d(s,\tilde s);\label{EQNa}\end{equation}
such a connection always exists \cite{BGb} and is
said to be a \it compatible unitary connection\rm. We note that equation
(\ref{EQNa}) does not determine $\nabla$ uniquely; there are
many compatible unitary connections. 


Use the inward geodesic flow
to identify a neighborhood of the boundary with the collar $\partial
M\times[0,\epsilon)$ for some $\epsilon>0$; if $y\in\partial M$, then the
curves $y(t)=(y,t)$ are unit speed geodesics perpendicular to the boundary. If
$y=(y^1,...,y^{m-1})$ are local coordinates on $\partial M$, let $x=(y,x^m)$ be
local coordinates on the collar where $x^m$ is the geodesic distance to the
boundary. Let $\partial_\mu:=\frac{\partial}{\partial x^\mu}$; $\partial_m$ is
the inward geodesic normal vector field on the collar. Let $\nabla_\mu$ be
covariant differentiation with respect to $\partial_\mu$. We may decompose 
$$P=\sum_{1\le\mu\le m}\gamma^\mu\nabla_\mu+\psi$$ 
where $\psi$ is a $0^{th}$ order operator. Since we do \it not \rm assume that
the structures are product near the boundary, the connection $1$ form of
$\nabla$, the leading symbol $\gamma$, and the endomorphism
$\psi$ can depend on the normal variable. Relative to a local frame on the
collar which is parallel along the normal geodesic rays which are perpendicular
to the boundary, we have $\nabla_m=\partial_m$. We set
$x^m=0$ to define a tangential operator
$$B_0(y):=\gamma^m(y,0)^{-1}\biggl(\
\sum_{\alpha<m}\gamma^\alpha(y,0)\nabla_\alpha+\psi(y,0)
   \biggr)\text{ on }C^\infty\biggl(E_1\big|_{\partial M}\biggr).$$
Let
$\Theta$ be an auxiliary self-adjoint endomorphism of $E_1|_{\partial M}$.
Let
$$A:= {\frac 1 2}\biggl(B_0+B_0^*\biggr)+\Theta\text{ on }
   C^\infty\biggl(E_1\big|_{\partial M}\biggr).$$  Here, the adjoint of $B_0$ is
taken with respect to the structures on the boundary. The operator $A$ is a
self-adjoint operator of Dirac type on
$C^\infty(E_1|_{\partial M})$. Let the boundary operator
${\mathcal B}$, which we will use to define the boundary conditions for the
operator $P$, be orthogonal projection on the span of the eigenspaces
for the non-negative spectrum of
$A$. Denote the realization of $P$ and the associated self-adjoint operator of
Laplace type by
$$P_{\mathcal B}\text{ and }D_{1,{\mathcal B}}:
    =(P_{\mathcal B})^*P_{\mathcal B}.$$

Let $F\in C^\infty(E_1)$ be an auxiliary function we use to localize the
problem; $F$ is called the \it smearing function. \rm Results of Grubb and Seeley
\cite{GSa, GSb, GSc} show that there is an asymptotic series as $t\downarrow0$ of the
form:
\begin{equation}\operatorname{Tr}_{L^2}(Fe^{-tD_{1,{\mathcal B}}})\sim
 \sum_{0\le k\le m-1}a_k(F,D_1,{\mathcal B})t^{(k-m)/2}+O(t^{-1/8}).
 \label{EQNb}\end{equation}
(There is a complete asymptotic series with log terms, but we shall not
need this fact as we shall only be interested in the first few terms in the
series).

The coefficients $a_k$ in equation (\ref{EQNb}) are locally computable. We shall
determine $a_0$, $a_1$, and $a_2$ in this paper. Our purpose is at least partly
expository, so we will present several different techniques which yield information
about these asymptotic coefficients.

We adopt the following notational conventions. Roman
indices $i$ and $j$ will range from $1$ to $m$ and index a local orthonormal frame for $TM$;
Greek indices will index a local coordinate frame.  Near the boundary, we
choose the frame so that $e_m$ is the inward unit geodesic normal vector; we
let indices $a$ and $b$ range from $1$ through $m-1$ and index the
corresponding frame for the tangent bundle of the boundary. We adopt the
Einstein convention and sum over repeated indices. We let `;' denote
multiple covariant differentiation of the tensors involved. Decompose
$$D_1=-(g^{\mu\nu}\partial_\mu\partial_\nu+a^\mu\partial_\mu+b).$$ 
Let $\Gamma$ be the Christoffel symbols of the Levi-Civita connection on
$M$. There is a canonical
connection ${}^D\nabla$ on the bundle $E_1$ and there is a canonical endomorphism
$E$ of the bundle $E_1$ so that 
$$D_1=-\{\Trace({}^D\nabla^2)+E\};$$
we refer to \cite{Gb} for details. If $E_1=E_2$ and if $P=P^*$, then
${}^D\nabla$ is a unitary connection; however ${}^D\nabla\gamma$ need not vanish in
general so ${}^D\nabla$ is not in general a compatible connection. Let
$\omega$ be the connection
$1$ form of
${}^D\nabla$. We have
\begin{eqnarray}
&&\omega_\delta:= {\frac 1 2}g_{\nu\delta}(a^\nu+g^{\mu\sigma}
    \Gamma_{\mu\sigma}{}^\nu)\text{ and}\nonumber\\
&&E:=b-g^{\nu\mu}(\partial_\nu\omega_\mu
   +\omega_\nu\omega_\mu-\omega_\sigma\Gamma_{\nu\mu}{}^\sigma).\label{EQNc}
   \end{eqnarray}
Decompose $P=\gamma_i\nabla_i+\psi$ and let $\hat\psi:=\gamma_m^{-1}\psi$. Let
${\mathcal R}$ be the scalar curvature of the metric on $M$. The main result of this
paper is the following:

\begin{theorem}\label{THM1} We have\begin{enumerate}
\item $a_0(F,D_1,{\mathcal B}) =  (4\pi )^{-m/2} \int_{M}\Trace(F)$.
\smallbreak\item 
      $a_1(F,D_1,{\mathcal B}) =  \frac{1}{4}
    \left[ \frac{\Gamma(\frac{m}{2})}
     {\Gamma(\frac{1}{2})\Gamma(\frac{m+1}{2})}-1\right](4\pi)^{-(m-1)/2}
    \int_{\partial M}\Trace(F)$.
\smallbreak\item $a_2(F,D_1,{\mathcal B})
     =(4\pi)^{-m/2}\int_M\frac{1}{6}\Trace
         \biggl\{F({\mathcal R}+6E)\biggr\}$\newline$
  \phantom{a}\qquad\qquad+(4\pi)^{-m/2}\int_{\partial M}\Trace\biggl\{
     \frac{1}{2}\left[\hat\psi+\hat\psi^*\right]F+
     \frac 1 3 \left[1- 
     \frac{3\Gamma(\frac{1}{2})\Gamma (\frac m 2)}{4\Gamma (\frac{m+1}2)} 
\right]L_{aa} F 
      $\newline$
      \phantom{a}\qquad\qquad\qquad\qquad
      -\frac{m-1}{2(m-2)} \left[ 1-
      \frac{\Gamma(\frac{1}{2})\Gamma (\frac m 2)}{2\Gamma
(\frac{m+1}2)}\right]F_{;m}
     \biggr\}.$
\end{enumerate}\end{theorem}

If $k<m$, then there exist locally computable invariants $a_k^M$ and $a_k^{\partial M}$
so that
\begin{eqnarray*}
 a_k(F,D_1,{\mathcal B})
    &=&(4\pi)^{-m/2}\int_M\Trace(a_k^M(F,D_1,x))\\
   &&+(4\pi)^{-m/2}\int_{\partial M}\Trace(a_k^{\partial
M}(F,D_1,{\mathcal B},y)).
\end{eqnarray*}
We have included a normalizing factor of $(4\pi)^{-m/2}$ to simplify
the formulas for the local invariants $a_k^M$ and $a_k^{\partial M}$.
We use dimensional analysis to see that the invariants $a_k^M$ 
are homogeneous of weight
$k$ and the invariants $a_k^{\partial M}$ are homogeneous of weight $k-1$ in
the jets of the symbols of $P$ and $P^*$. The invariants $a_{2j+1}^M$ vanish. We use
Theorem  4.1.6 \cite{Gc} to see that:
\begin{equation}a_0^M(F,D_1,x)=\Trace\{F\}\text{ and }
  a_2^M(F,D_1,x)=\Trace\{F( {\frac{1}{6}}{\mathcal R}+E)\}.
  \label{EQNx}\end{equation}

The bundles $E_1$ and $E_2$ are distinct; we must use $\gamma_m$ to
identify $E_1$ and $E_2$ near the boundary. This observation reduces the number of
invariants which are homogeneous of weight 1; for example, $\Trace(\psi)$ is not
invariantly defined. There are universal constants so that
\begin{eqnarray*}
  &&a_1^{\partial M}(F,D_1,{\mathcal B})=b_1(m)\Trace(F)\\
    &&a_2^{\partial M}(F,D_1,{\mathcal B},y)
     =c_0(m)F(\hat\psi+\hat\psi^*)+
    c_1(m)F(\hat\psi-\hat\psi^*)\\&&\qquad\qquad\qquad\qquad
   +c_2(m)F\Theta+c_3(m)FL_{aa}+c_4(m)F_{;m}.\end{eqnarray*}
In contrast to the situation when the boundary operator ${\mathcal B}$ is local, the
constants exhibit non-trivial dependence upon the dimension.

We will use three different methodologies to compute the unknown
coefficients. In Section \ref{SEC2}, we use results of Grubb and Seeley
\cite{GSc} for structures that are product near the boundary to compute the
constants $b_1(m)$ and $c_0(m)$; see Lemma
\ref{THM2} for details. In Section \ref{SEC3}, we use
functorial properties of these invariants to determine the coefficients
$c_0(m)$, $c_1(m)$, and $c_2(m)$; see Lemma
\ref{THM3} for details.  In Section \ref{SEC4}, we use computations on
the ball to determine the coefficients $b_1(m)$, $c_3(m)$, and
$c_4(m)$; see Lemma \ref{THM4} for details. As a check on our
methods, we give two different derivations of the relation
\begin{equation}(m-2)c_4(m)+(m-1)c_3(m)=- {\frac{m-1}{6}}
\end{equation}
in Sections \ref{SEC3} and \ref{SEC4}. Our purpose in this paper is partly
pedagogical; we wish to illustrate different methodologies which can be used to
study these invariants.

\bigbreak\section{Product Formulas\label{SEC2}} We use results of Grubb and
Seeley \cite{GSc} to show:

\begin{lemma}\label{THM2}\ \begin{enumerate}
\item We have  $b_1(m)= \frac{1}{4}
     \left[ \frac{\Gamma(\frac{m}{2})}
     {\Gamma(\frac{1}{2})\Gamma(\frac{m+1}{2})}-1\right](4\pi)^{-(m-1)/2}$.
\smallbreak\item We have  $c_0(m)=\frac{1}{2}$.
\end{enumerate}
\end{lemma}

\noindent\bf Proof of Lemma \ref{THM2}: \rm Suppose that $P=\gamma_m(\nabla_m+B)$ where
$B$ is a self-adjoint tangential operator of Dirac type with coefficients which are
independent of the normal variable; we take $\Theta=0$ so $A=B$. In this setting, we
say the \it structures are product near the boundary. \rm Let $\tilde D_1$ be the
associated operator of Laplace type on the double. We ignore the effect of the
$0$ spectrum and define:
\begin{eqnarray*}
\eta(s,B):=\Trace_{L^2}(B(B^2)^{-s-1}),&&
   \zeta(2s,B):=\Trace_{L^2}((B^2)^{-s})\\
\zeta(s,\tilde D_1):=\Trace_{L^2}(\tilde D_1^{-s}),&&
   \zeta(s,D_1):=\Trace_{L^2}(D_1^{-s}).
\end{eqnarray*}
We refer to
Theorem 2.1
\cite{GSc} for the proof that:
\begin{eqnarray}
   &&\Gamma(s)\zeta(s,D_1)=R(s)+
    \Gamma(s)\biggl\{{ \frac{1}{2}}\zeta(s,\tilde D_1)
   + \frac{1}{4}\bigl(
    \frac{\Gamma(s+{ \frac{1}{2}})}
   {\Gamma({ \frac{1}{2}})\Gamma(s+1)}-1\bigr)\zeta(s,B^2)\nonumber\\
  &&\qquad\qquad\qquad\qquad- \frac{1}{4} 
   \frac{\Gamma(s+{ \frac{1}{2}})}
   {\Gamma({ \frac{1}{2}})\Gamma(s+1)}\eta(2s,B)\biggr\}
     \label{EQNe}
\end{eqnarray}
where the remainder $R$ is regular away from $s=0$.
Expand
$$\Trace_{L^2}(Be^{-tB^2})\sim \sum_k 
a_k^\eta(B)t^{(-(m-1)+k-1)/2}.$$
The invariants $a_{2j}^\eta$ vanish. Formulas of Branson and
Gilkey \cite{BGb} show that
\begin{equation}a_1^\eta(B)
    =-(4\pi)^{-(m-1)/2}(m-2) \int_{\partial M}\Trace(\hat\psi).\label{EQNy}
\end{equation}
One can use the Mellin transformation to relate the asymptotics of the heat
equation to the pole structure of the eta and zeta
functions. Let $N$ be a manifold of dimension $n$, let $D_N$ be an operator of Laplace
type on $N$, and let $Q_N$ be an operator of Dirac type on $N$. If the boundary
of $N$ is non-empty, impose spectral boundary conditions. We then have, see for example
Theorem 1.12.2 \cite{Gc},
\begin{eqnarray}
&&a_k(D_N)=\text{Res}_{s=\frac{n-k}{2}}\Gamma(s)\zeta(s,D_N)\text{ and }\nn\\
&&a_k^\eta(Q_N)=\text{Res}_{s=\frac{n-k-1}2}\Gamma(s)\eta(2s-1,Q_N).
\label{EQNw}\end{eqnarray}
 The following
identities now follow from equations (\ref{EQNe}) and (\ref{EQNw}):
\begin{eqnarray}
&&a_n(1,D_1,{\mathcal B})= \frac{1}{2}a_n(1,\tilde D_1)
   - \frac{1}{2(m-n)\Gamma( \frac{1}{2})}a_{n-1}^\eta(1,B)\text{ if
}n\equiv0
   \text{ mod }2,\label{EQNf}\\
&&a_n(1,D_1,{\mathcal B})=
    \frac{1}{4}\bigl(
    \frac{\Gamma({ \frac{m-n+1}{2}})}
  {\Gamma({ \frac{1}{2}})\Gamma({ \frac{m-n+2}{2}})}-1\bigr)
    a_{n-1}(1,B^2)\text{ if }n\equiv1
   \text{ mod }2.
\label{EQNg}\end{eqnarray}
We prove the first assertion of Lemma \ref{THM2} by using equations (\ref{EQNx}) and
(\ref{EQNg}) to compute:
\begin{eqnarray*}
&&\phantom{=}(4\pi)^{-m/2}a_1^{\partial M}(1,D_1,{\mathcal B},y)\\
&&=\frac{1}{4}
     \biggl( \frac{\Gamma(\frac{m}{2})}
    {\Gamma(\frac{1}{2})\Gamma(\frac{m+1}{2})}-1\biggr)(4\pi)^{-(m-1)/2}
   a_0(1,B^2,y)\\
&&=(4\pi)^{-(m-1)/2}\frac{1}{4}
     \biggl( \frac{\Gamma(\frac{m}{2})}
     {\Gamma(\frac{1}{2})\Gamma(\frac{m+1}{2})}-1\biggr)
   \Trace(1).\end{eqnarray*}
We prove the second assertion of Lemma \ref{THM2} by using equations (\ref{EQNy}) and
(\ref{EQNf}) to compute:
\begin{eqnarray*}
&&\phantom{=}(4\pi)^{-m/2}a_2^{\partial M}(1,D_1,{\mathcal B},y)\\
&&=-\frac{1}{2(m-2)\Gamma( {\frac{1}{2}})}(4\pi)^{-(m-1)/2}a_1^\eta(1,B,y)\\
&& =(4\pi)^{-m/2}\Trace(\hat\psi).\ \qedbox\end{eqnarray*}

\bigbreak\section{Functorial Method\label{SEC3}} The invariants $a_k$ have many
functorial properties. We use these properties to establish the following result.

\begin{lemma}\label{THM3}\ \begin{enumerate}
\item We have $c_2(m)=0$.
\item We have $c_0(m)= {\frac 1 2}$.
\item We have $c_1(m)=0$.
\item We have $(m-2)c_4(m)+(m-1)c_3(m)=- {\frac{m-1}{6}}$.
\end{enumerate}\end{lemma}

\noindent\bf Proof of Lemma \ref{THM3}: \rm We take $E_1=E_2$ and let $P$ be a formally
self-adjoint operator of Dirac type on $C^\infty(E_1)$. Let
$\Theta(\varepsilon):=\Theta+\varepsilon$ where $\varepsilon$ is a real parameter.
For generic values of $\varepsilon$, $\ker(A(\varepsilon))$ is trivial. For such
a value, the boundary conditions determined by the boundary
operator ${\mathcal B}(\varepsilon)$ are locally constant and thus $a_2$ is
independent of $\varepsilon$. The first assertion of Lemma \ref{THM3} now follows; this
implies that $a_2(\cdot)$ is not sensitive to the particular boundary condition
chosen among the family we are considering.

For the remainder of the proof of Lemma \ref{THM3}, let 
$\pbgT:=S^1\times...\times S^1$ be
the
$m-1$ dimensional torus and let $M:=\pbgT \times[0,1]$; we give
$\pbgT $ and $M$ the canonical product flat metrics and let $(x_1,...,x_m)$ be the
usual parameters. We identify $\pbgT $ with $\pbgT \times\{0\}$ in $M$. Let $L$ be
the line bundle defined by a non-trivial ${\mathbb Z}_2$ valued representation
of the fundamental group of $\pbgT$ and let $V$ be the trivial bundle of
dimension $2^m$ with coefficients in $L$. Let
$\gamma_i$ be \it real \rm skew-adjoint matrices acting on
$V$ which satisfy the Clifford
commutation relations 
$$\gamma_i\gamma_j+\gamma_j\gamma_i=-2\delta_{ij}.$$
The twisting defined by $L$ ensures that the kernel of the associated tangential
operator $B$ is trivial. Let $\nabla_i:=\partial_i$ define a compatible unitary
connection on
$C^\infty(V)$. Let $\Delta_0:=(\gamma_i\partial_i)^2=-\partial_i^2$ be the
associated operator of Laplace type on $C^\infty(V)$. Let
$f=f(x_m)$ be a smooth real valued function on $M$ which vanishes identically
near
$x_m=1$.

We use the index theorem to evaluate the coefficient $c_0(m)$. Let
\begin{eqnarray*}
  &&Q:=\gamma_i\partial_i+f\gamma_m,\qquad Q^*:=\gamma_i\partial_i-f\gamma_m,\\
 &&D_1:=Q^*Q=\Delta-2\gamma_m\gamma_af\partial_a+f^2-f_{;m},\\
 &&D_2:=QQ^*=\Delta+2\gamma_m\gamma_af\partial_a+f^2+f_{;m}.\end{eqnarray*}
We use equation (\ref{EQNc}) to compute:
\begin{eqnarray*}
   &&\omega_{1,a}=\gamma_m\gamma_af,\phantom{-}\ \omega_{1,m}=0,\
      E_1=f_{;m}+(m-2)f^2,\\
   &&\omega_{2,a}=-\gamma_m\gamma_af,\ \omega_{2,m}=0,\
      E_2=-f_{;m}+(m-2)f^2.\end{eqnarray*}
We have $\hat\psi_1=f$ and $\hat\psi_2=-f$. The local formula for the index shows
that the super trace vanishes for $n\ne m$. Since $m\ge3$, we have
\begin{eqnarray*}
  &&  0=a_2(1,D_1,{\mathcal B})-a_2(1,D_2,{\mathcal B})\\
  &&  \phantom{0}=\dim(V)\bigl\{2\int_Mf_{;m}
  +4c_0(m)\int_\pbgT f\bigr\}\\
  && \phantom{0}=\dim(V)(-2+4c_0(m))
 \int_\pbgT f.\end{eqnarray*}
This shows that $c_0(m)= {\frac 1 2}$ and proves the second
assertion of Lemma \ref{THM3}; note that this value agrees with the result obtained
previously in the proof of Lemma \ref{THM2}.

Next we study the coefficient $c_1$. Let
$$P_f:=\gamma_i\partial_i+\sqrt{-1}f\gamma_m;$$
this operator is formally self-adjoint.
Note that $A_f=-\gamma_m\gamma_a\nabla_a$ so the boundary operator ${\mathcal B}$ is
independent of the function $f$. We expand
$$P_f^2=\Delta_0-2\sqrt{-1}f\partial_m-\sqrt{-1}f_{;m}+f^2.$$
Since the $\gamma_i$ were real, complex conjugation preserves the boundary conditions
and intertwines $P_f^2$ with $P_{-f}^2$. Thus $P_f^2$ and $P_{-f}^2$ are
isospectral so
\begin{equation}a_2(1,P_f^2,{\mathcal B})=a_2(1,P_{-f}^2,{\mathcal B}).
\label{EQNh}
\end{equation}
We use equation (\ref{EQNc}) to see the interior
integrand vanishes by computing:
$$\omega_m(f)=\sqrt{-1}f\text{ and }E(f)=0.$$
Since $\hat\psi_f=\sqrt{-1}f$, we may use equation (\ref{EQNh}) to show
$c_1=0$ by computing:
$$0=a_2(1,P_f^2,{\mathcal B})-a_2(1,P_{-f}^2,{\mathcal B})
      =4\sqrt{-1}c_1(m) \int_\pbgT f.$$

We can also show $c_1(m)=0$ using gauge invariance. Let
$\tilde P=e^{-\sqrt{-1}f}P_0e^{\sqrt{-1}f}$ be defined by a
global unitary change of gauge. We have
$$\psi_f=\psi+\sqrt{-1}f_{;i}\gamma_i\text{ so }
  \hat\psi_f=\hat\psi+\sqrt{-1}f_{;i}\gamma_m^{-1}\gamma_i.$$
Thus $\Trace(\hat\psi_f-\hat\psi_f^*)=-2\sqrt{-1}f_{;m}\dim(V)$.
As $a_2$ is gauge invariant, $c_1(m)=0$.

To prove the final assertion of Lemma \ref{THM3}, we shall
need the following technical result; we postpone the proof until the
end of this section.

\begin{sublemma}\label{THM3a} Let
$ds^2(\varepsilon)=e^{2\varepsilon f}ds^2$
   and let $P(\varepsilon):=e^{-\varepsilon f}P$. There exists a compatible
   family of unitary connections $\nabla(\varepsilon)$ so that
$$ \psi(\varepsilon)
   =e^{-\varepsilon f}\biggl\{\psi(0)-\frac{m-1}2\varepsilon\biggr\}f_{;i}\gamma_i.$$
\end{sublemma}

We use Sublemma \ref{THM3a} to complete the proof of Lemma \ref{THM3}. Let
$P_0:=\gamma_i\partial_i$ and let
\begin{eqnarray*}
  &&ds^2(\varepsilon)=e^{2\varepsilon f}ds^2,\qquad
    P(\varepsilon):=e^{-\alpha(m)\varepsilon f}P_0e^{-\beta(m)\varepsilon
   f},\\
  &&dvol(\varepsilon)=e^{m\varepsilon f}dvol,\quad
   P(\varepsilon)^*=e^{(-\beta(m)-m)\varepsilon
   f}P_0^*e^{(m-\alpha(m))\varepsilon f}.\end{eqnarray*}
We fix the inner product on $V$. Let
$\alpha(m)= {\frac{1+m}{2}}$ and let
$\beta(m):= {\frac{1-m}{2}}$.
Then $\alpha(m)+\beta(m)=1$ and $\alpha(m)=\beta(m)+m$ so the metric
determined by the leading symbol of $P(\varepsilon)$ is $ds^2(\varepsilon)$ and
$P(\varepsilon)$ is formally self-adjoint. We assume that $f=f(x_m)$ vanishes on the boundary
of $M$ and that $f$ vanishes identically near $x_m=1$. Thus the leading symbol of
$A(\varepsilon)$ is independent of
$\varepsilon$ so by defining
$\Theta(\varepsilon)$ appropriately, we can assume the boundary operator
${\mathcal B}(\varepsilon)$ is independent of $\varepsilon$. Let
$D(\varepsilon):=P(\varepsilon)^2$. Let
$\delta:={ {\frac{\partial}{\partial\varepsilon}}}|_{\varepsilon=0}$
We compute the variation:
\begin{eqnarray*}
  &&\delta\Trace_{L^2}(e^{-tD(\varepsilon)})=
     -t\Trace_{L^2}(\{\delta
      D(\varepsilon)\}e^{-t\Delta_0})\\
     &&\qquad=-2t\Trace_{L^2}(\{\delta P(\varepsilon)\}P_0e^{-t\Delta_0})
    =2t\Trace_{L^2}(f\Delta_0e^{-t\Delta_0})\\
    &&\qquad=-2t\partial_t\Trace_{L^2}(fe^{-t\Delta_0}).\end{eqnarray*}
We equate coefficients in the asymptotic expansions to see
\begin{equation}
   \delta a_2(1,D(\varepsilon),{\mathcal B})=
  (m-2)a_2(f,\Delta_0,{\mathcal B}).\label{EQNi}\end{equation}

If $\varepsilon=0$, then $D=\Delta_0$ and $a_2^M=0$. We are interested in
the coefficient of $\varepsilon f_{;mm}$.
Since $\omega(0)=0$, since $\Gamma(0)=0$, and since $E(0)=0$, we may compute
\begin{eqnarray*}
    &&b(D(\varepsilon))=-\beta(m)\varepsilon f_{;mm}+O(\varepsilon^2)\\
    &&a^m(D(\varepsilon))=(m-2\alpha(m)-2\beta(m))\varepsilon f_{;m}\\
    &&\omega_m(D(\varepsilon))=
          {\frac{m-2\alpha(m)-2\beta(m)}{2}}\varepsilon f_{;m}
          +\omega_m(e^{-2\varepsilon f}\Delta_0),\\
     &&E(D(\varepsilon))=b(\varepsilon)
              -\partial_m\omega_m(\varepsilon)
         +E(e^{-2\varepsilon f}\Delta_0)+O(\varepsilon^2)\\
     &&\phantom{E(D(\varepsilon))}=
          - {\frac{m-2\alpha(m)}{2}}\varepsilon
           f_{;mm}+E(e^{-2\varepsilon f}\Delta_0)+O(\varepsilon^2).\end{eqnarray*}
We use results of \cite{BGa} to see that
\begin{eqnarray*}
  &&\delta E(e^{-2\varepsilon f}\Delta_0)=-2fE+ {\frac 1
      2}(m-2)f_{;ii},\\
  &&\delta{\mathcal R}=-2f{\mathcal R}-2(m-1)f_{;ii},\text{ and}\\
  &&\delta L_{aa}=-fL_{aa}-(m-1)f_{;m}.\end{eqnarray*}
This permits us to compute the variation of the interior integral:
\begin{eqnarray}
   && \delta\int_M a_2^M(D(\varepsilon))
        =\biggl\{-\frac{2m-2}{6}+\frac{m-2}{2}-\frac{m-2\alpha(m)}{2}\biggr\}
         \int_M f_{;mm}\nonumber\\
   && \phantom{\delta \int_M a_2^M(D(\varepsilon))}
        =\biggl\{-\alpha(m)+{\frac{m+2}{3}}\biggr\}\int_\pbgT 
     f_{;m}.\label{EQNj}
\end{eqnarray}
We have 
$P(\varepsilon)=e^{-\varepsilon f}(P_0-\beta(m)\varepsilon\gamma_mf_{;m})$.
Since $\psi(P_0)=0$, Lemma \ref{THM3a} (2) implies that
$ \psi(e^{-\varepsilon f}P_0)=
   -\frac{m-1}2\varepsilon e^{-\varepsilon
f} f_{;i}\gamma_{i}$.  As $f$ vanishes on $\partial M$,
\begin{eqnarray}
   && \delta\hat\psi(P(\varepsilon))=-\biggl\{\beta(m)+\frac{m-1}2\biggr\}f_{;m}
      \text{ and}\nonumber\\
  && \delta\int_\pbgT  a_2^{\partial M}(D(\varepsilon),{\mathcal
B})
     =\biggl\{-(m-1)c_3(m)-\beta(m)-\frac{m-1}2\biggr\}\int_\pbgT 
f_{;m}.\label{EQNk}
\end{eqnarray}
Recall that $\alpha(m)+\beta(m)=1$ and that $a_2(\Delta_0)=0$. We use equations
(\ref{EQNi}), (\ref{EQNj}), and (\ref{EQNk}). The final assertion of Lemma
\ref{THM3} follows from the following equation:
$$(m-2)c_4(m)=-\alpha(m)+ {\frac{m+2}{3}}-(m-1)c_3(m)-\beta(m)
    -\frac{m-1}2.\
\qedbox$$

\noindent\bf Proof of Sublemma \ref{THM3a}: \rm We follow the
argument given in \cite{Gc}.
Let $M$ be an arbitrary Riemannian manifold and let
$\gamma$ be a Clifford module structure on a vector bundle over $M$. Let
$\partial_\mu$ and $dx^\mu$ be local coordinate frames for the tangent and
cotangent bundles. We define:
$$\gamma^{\mu}(\varepsilon):=e^{-f\varepsilon}\gamma^{\mu}\qquad
  \gamma_{\mu} (\varepsilon):=e^{f\varepsilon}\gamma_{\mu} \qquad
   \theta_{\mu} := {\frac\varepsilon4}\biggl\{2f_{;\nu}\gamma^{\nu}\gamma_{\mu}
   +cf_{;\mu}\biggr\}.$$
Let $\nabla(\varepsilon):=\nabla+\theta$.
Since the original connection is compatible, we have
$$0=\nabla_{\mu} (\gamma)(dx^{\nu})
  =\partial_{\mu}(\gamma^{\nu})+[\omega_{\mu} ,\gamma^{\nu}]
  +\Gamma_{\mu \sigma}{}^{\nu}\gamma^\sigma(0).$$
Let ${\mathcal E}_{\mu }(\varepsilon):=\nabla_{\mu} (\varepsilon)\gamma(\varepsilon)$
and let ${\mathcal E}_{\mu} ^{\nu}:={\mathcal E}_{\mu} (dx^{\nu})$. We then have
\begin{eqnarray*}
&&{\mathcal E}_{\mu} ^{\nu}(\varepsilon)=-\varepsilon f_{;\mu}\gamma^{\nu}(\varepsilon)
    +[\theta_{\mu},\gamma^{\nu}](\varepsilon)
     +(\Gamma_{\mu \sigma}{}^{\nu}(\varepsilon )
    -\Gamma_{\mu \sigma}{}^{\nu}(0))\gamma^{\sigma}(\varepsilon).
\end{eqnarray*}
Fix $x_{0}\in M$ and choose the local coordinates near $x_0$ so
$g_{\mu\nu}(x_{0})=\delta_{\mu\nu}$. Then
\begin{eqnarray*}
 &&\Gamma_{\mu\nu}{}^{\sigma}(x_{0})= { {\frac 1 2}}
   g^{\sigma\tau}(\partial_{\nu}g_{\mu\tau}+\partial_{\mu} 
   g_{\nu\tau}-\partial_\tau g_{\mu\nu})(x_{0})\\
 &&(\Gamma_{\mu \sigma}{}^{\nu}(\varepsilon)-\Gamma_{\mu \sigma}{}^{\nu})(x_{0})
   =\varepsilon(\delta_{\mu\nu}f_{;\sigma}+\delta_{\nu\sigma}f_{;\mu}
   -\delta_{\mu \sigma}f_{;\nu})(x_{0})\\
  &&\theta_{\mu}(x_{0})= {\frac\varepsilon4}
   \biggl\{2f_{;\sigma}\gamma_{\sigma}\gamma_{\mu}+cf_{;\mu})\biggr\}(x_{0}).
\end{eqnarray*}
We must show
${\mathcal E}_{\mu}^{\nu}(x_{0})=0$. If $\mu\not=\nu$, we compute:
\begin{eqnarray*}
  &&{\mathcal E}_{\mu}^{\nu}(x_{0})
  =e^{-\varepsilon f}\{-\varepsilon f_{;\mu}\gamma_{\nu}+ { {\frac 1 2}}\varepsilon
    f_{;\sigma}[\gamma_{\sigma}\gamma_{\mu},\gamma_{\nu}]+\varepsilon
(-f_{;\nu}\gamma_{\mu}+f_{;\mu}\gamma_{\nu})\}
    (x_{0})\\
   &&\phantom{{\mathcal E}_{\mu}^{\nu}(x_{0})}=
     e^{-\varepsilon f}\{-\varepsilon f_{;\mu}\gamma_{\nu}+\varepsilon
    f_{;\sigma}\gamma_{\mu}\delta_{\nu\sigma}\}(x_{0})=0.\end{eqnarray*}
If $\mu=\nu$, then (don't sum over $\mu$):
\begin{eqnarray*}
   && {\mathcal E}_{\mu}^{\mu}(x_0)=e^{-\varepsilon f}\biggl\{-\varepsilon
     f_{;\mu}\gamma^{\mu}+[\theta_{\mu},\gamma^{\mu}]
    +\sum_{\sigma}(\Gamma_{\mu \sigma}{}^{\mu}(\varepsilon
       )-\Gamma_{\mu \sigma}{}^{\mu}(0))\gamma_{\sigma})\biggr\}(x_0)\\
    && \phantom{{\mathcal E}_{\mu}^{\mu}}=-e^{-\varepsilon f}\biggl\{\varepsilon
      f_{;\mu}\gamma_{\mu }
       -\varepsilon\sum_{\sigma\not=\mu
     }f_{;\sigma}\gamma_{\sigma}
    +\varepsilon\sum_{\sigma}f_{\sigma}\gamma_{\sigma}\biggr\}(x_0)=0.\end{eqnarray*}
This shows that $\nabla(\varepsilon)$ is compatible. Let $c=2$.
Then $\theta_\mu+\theta_\mu^*=0$ and $\nabla(\varepsilon)$ is unitary. 
We complete the proof of Sublemma \ref{THM3a} by computing
\begin{eqnarray*}
    &&\psi (\varepsilon )=e^{-\varepsilon
    f}\biggl\{\psi-{ {\frac1{4}}}\varepsilon
    (2f_{;\nu}\gamma^{\mu}\gamma^{\nu}\gamma_{\mu}
     +cf_{;\mu}\gamma^{\mu})\biggr\}\\
    &&\phantom{\psi(\varepsilon)}=e^{-\varepsilon f}\biggl\{\psi -
       { {\frac1{4}}}\varepsilon f_{;\mu}\gamma^{\mu}(2m-4+c)\biggr\}.\ \qedbox
\end{eqnarray*}


\bigbreak\section{Computations on the Disk\label{SEC4}} 

We perform computations on the disk to prove:
\begin{lemma}\label{THM4}\ 
\begin{enumerate}
\item We have $b_1(m)= \frac{1}{4}
    \left[ \frac{\Gamma(\frac{m}{2})}
     {\Gamma(\frac{1}{2})\Gamma(\frac{m+1}{2})}-1\right](4\pi)^{-(m-1)/2}$.
\smallbreak\item We have $c_3 (m) = \frac 1 3 \left[1-
\frac{3\Gamma({\frac{1}{2}})\Gamma ({\frac m 2})}{4\Gamma (\frac{m+1}{2})}
\right]$.
\smallbreak\item We have $c_4 (m) = -\frac{m-1}{2(m-2)} \left[ 1-\frac 1 2 \Gamma(
\frac{1}{2})
\frac{\Gamma (\frac m 2)}{\Gamma (\frac{m+1}{2})}\right]$.
\smallbreak\item We have $(m-2)c_4 (m) +(m-1) c_3 (m) = -\frac {m-1} 6$ .
\end{enumerate}
\end{lemma}

\noindent\bf Proof: \rm
Let $M$ be the unit ball in ${\mathbb R}^m$ with the usual metric.
If $r\in[0,1]$ is the radial normal coordinate and if $d\Sigma^2$ is
the usual metric on the unit sphere $S^{m-1}$, then
$ds^2=dr^2+r^2d\Sigma^2$.
The inward unit normal on the boundary is $-\partial_r$. The only
nonvanishing components of the Christoffel symbols are
\begin{eqnarray}
\Gamma_{abc}= \frac 1 r \tilde{\Gamma}_{abc}\text{ and }
\Gamma_{abm}= \frac 1 r \delta_{ab};\nn
\end{eqnarray}
the second fundamental form is given by $\Gamma_{abm}=L_{ab}$. Here
$\tilde{\Gamma}_{abc}$ are the Christoffel symbols associated with the
metric $d\Sigma^2$ on the sphere $S^{m-1}$ and tilde will always refer to this metric.

The spin representation $\gamma$
is an irreducible representation of the Clifford algebra; we refer to
\cite{ABS} for details. Let $P=\gamma^\nu\partial_\nu$ be the Dirac operator on the
ball; we take the flat connection $\nabla$ and set
$\psi=0$. We suppose $m$ even (there is a corresponding
decomposition for $m$ odd) to find a local decomposition:
\begin{eqnarray}
&&\gamma^a_{(m)}=\left(
   \begin{array}{cc}
               0 &  \sqrt{-1}\cdot \gamma^a_{(m-1)}    \\
      -\sqrt{-1}\cdot \gamma^a_{(m-1)}    &     0
    \end{array}    \right)\text{ and }\nn\\
\quad
&&\gamma^m_{(m)} = \left(
     \begin{array}{cc}
         0       &    \sqrt{-1}\cdot 1_{m-1}   \\
    \sqrt{-1}\cdot 1_{m-1}\quad\    &      0
    \end{array}   \right)   .\nonumber
\end{eqnarray}
We stress that $\gamma^j_{(m)}$ are the $\gamma$-matrices projected
along some vielbein system. Decompose $\nabla_j = e_j + \omega_j$ where
$\omega_j=\frac 1 4 \Gamma_{jkl} \gamma_k \gamma_l$ is the connection $1$ form
of the spin connection. Note that
$$
\nabla_a = \frac 1 r\left( \left(
        \bea {cc}
        \tilde{\nabla}_a & 0 \\
         0 & \tilde{\nabla}_a
         \eea  \right)  +\frac 1 2 \delta_{ab}
        (-\gamma^m_{(m)} \gamma^b_{(m)})\right).
$$
Let $\tilde P$ the Dirac operator on the sphere. We have:
\begin{eqnarray}
&&P=\left(\frac{\partial}{\partial x_m}-\frac{m-1}{2r} \right) \gamma_{(m)}^m
         +\frac 1 r \left(
\begin{array}{cc}
       0   & \sqrt{-1} \tilde P \\
    -\sqrt{-1} \tilde P  & 0
\end{array}   \right).\nn
\end{eqnarray}
Let $d_s$ be the dimension of the spin bundle on the disk; $d_s=2^{m/2}$ if $m$ is even.
The spinor modes ${\mathcal Z} _\pm ^{(n)}$ on the sphere are discussed in
\cite{roberto}. We have 
\begin{eqnarray*}
&&\tilde P {\mathcal Z} _\pm ^{(n)} (\Omega )=  \pm \left( n+\frac{m-1} 2
\right)
             {\mathcal Z} _\pm ^{(n)} (\Omega )\text{ for }n=0,1,...;\\
&&  d_n(m):=\dim {\mathcal Z} _\pm ^{(n)}= \frac 1 2 d_s \left(
    \bea {c}
       m+n-2 \\
        n
      \eea \right) .
\end{eqnarray*}

Let $J_{\nu} (z)$ be the Bessel functions. These satisfy the differential equation and
functional relations \cite{grad}:
\begin{eqnarray*}
&& \frac{d^2 J_{\nu} (z)} {dz^2} +\frac 1 z \frac{dJ_{\nu} (z)} {dz} +
\left( 1-\frac{\nu^2}{z^2} \right) J_{\nu }(z) =0,\\
&&  z\frac d {dz} J_{\nu} (z) +\nu J_{\nu} (z)=zJ_{\nu -1} (z),\text{ and} \\
&&  z \frac d {dz} J_{\nu } (z) -\nu J_{\nu} (z)=-z J_{\nu +1}
(z).\end{eqnarray*} 
Let  $P\varphi_\pm = \pm \lambda \varphi_\pm$ be an eigen function
of $P$. Modulo a suitable radial normalizing constant $C$, we may express:
\begin{eqnarray}
\varphi_{\pm}^{(+)}&=&{\frac{C}{r^{(d-1)/2}}} \left(
     \begin{array}{c}
        iJ_{n+m/2}(kr)
       \,Z^{(n)}_+(\Omega ),  \\
     \pm J_{n+m/2-1}(kr)\,Z^{(n)}_+(\Omega )
        \end{array}  \right) , \text{ and} \label{eq2.59}\\
\varphi_{\pm}^{(-)}&=&{\frac{C}{r^{(d-1)/2}}}\left(
     \begin{array}{c}
     \pm J_{n+m/2-1}(kr)\,Z^{(n)}_
-(\Omega )  \\
   iJ_{n+m/2}(kr)\,Z^{(n)}_-(\Omega ) \end{array}
\right).
\label{solutions}
\end{eqnarray}
Let ${}^T\gamma^a_{(m)}:=-\gamma_{(m)}^m\gamma_{(m)}^a$ and let
${}^T\nabla_a:=\nabla_a- \frac 1 2 L_{ab}{}^T\gamma_{(m)}^b$.
Then ${}^T\nabla$ is a compatible unitary connection for the induced
Clifford modules structure ${}^T\gamma$; see \cite{G} for details.
We may express the tangential operator $B$ in the form:
\begin{eqnarray}
B &=&-\gamma^m_{(m)} \gamma^a_{(m)}\nabla_a  = {^T\gamma^a_{(m)}} \nabla_a =
{^T\gamma^a_{(m)}}
\left({^T\nabla_a} + \frac 1 2 L_{ab} {^T\gamma_{(m)}^b}\right)
\nonumber\\
&=& \left(
\begin{array}{cc}
       -\tilde P -\frac {m-1} 2 & 0 \\
   0 & \tilde P -\frac {m-1} 2
 \end{array} \right).\nn
\end{eqnarray}
Thus in particular $B=B^*$. We take $\Theta = \frac{m-1} 2 \,\, 1_m$. We then have:
\begin{eqnarray}
A=\left(
\begin{array}{cc}
-\tilde P & 0 \\
0 & \tilde P
\end{array} \right). \nn
\end{eqnarray}
The eigenstates and eigenvalues of $A$ then are given by:
\beq
A\left(\bea {c}
          {\mathcal Z}_+^{(n)} \\
            {\mathcal Z}_-^{(n)}
        \eea \right) &=& -\left( n+\frac{m-1} 2 \right)
\left(\bea {c}
          {\mathcal Z}_+^{(n)} \\
            {\mathcal Z}_-^{(n)}
        \eea \right)\text{ and} \nn\\
A\left(\bea {c}
          {\mathcal Z}_-^{(n)} \\
            {\mathcal Z}_+^{(n)}
        \eea \right) &=&  \left( n+\frac{m-1} 2 \right)
\left(\bea {c}
          {\mathcal Z}_-^{(n)} \\
            {\mathcal Z}_+^{(n)}
        \eea \right)\text{ for }n=0,1,....\nn
\eeq
The boundary condition suppresses the
non-negative spectrum of $A$. We use equation (\ref{solutions}) to see that
the non-negative modes of $A$ are associated with the radial factor
$J_{n+\frac m 2 -1} (\lambda r)$. Hence the implicit eigenvalue
equation is
\beq
J_p (\lambda ) =0\text{ where }p=n+\frac m 2-1.\label{implicit}
\eeq
The first and third authors developed a method for calculating the associated
heat-kernel coefficients for smearing function $F=1$ in \cite{bek,cmp,jon};
they generalized this method to deal with $F=F(r)$ in
\cite{smear}. We summarize the essential results from these papers briefly;
in principal one could calculate any number of coefficients. 
We first suppose that
$F=1$. Instead of looking directly at the heat-kernel we will consider the
zeta-function $\zeta (s)$ of the operator $P^2$ and use 
the relationship provided by
equation (\ref{EQNw}) between the pole structure of the zeta function and the
asymptotics of the heat equation:
\beq
a_k =  \mbox{Res }_{s=\frac {m-k}2}\Gamma(s)\zeta (s).\label{ex1}
\eeq
Thus to compute $a_0$, $a_1$, and $a_2$, we must determine the residues of the
zeta-function $\zeta (s)$ at the values
$s=\frac m 2$, $s=\frac{m-1}2$, and $s=\frac m 2-1$. We use the eigenvalue equation
(\ref{implicit}) to express
\beq 
\zeta (s) = 4 \sum_{n=0}^\infty d_n (m)
\int_{\mathcal C} \frac{dk}{2\pi i} k^{-2s} \frac \partial {\partial k} \ln
J_p (k) ,
\label{ex2}
\eeq
where the contour ${\mathcal C}$ runs counterclockwise and encloses all the solutions
of (\ref{implicit}) which lie on the positive real axis. The factor of four
comes from the four types of solutions in (\ref{eq2.59}) and (\ref{solutions}).
As it stands, equation (\ref{ex2}) is well defined only for
$\Re s > m/2$, so the first task is to construct the analytical continuation to
the left. We define a modified zeta function
\beq 
\zeta ^{(n)} (s) =
\int_{\mathcal C} \frac{dk}{2\pi i} k^{-2s} \frac \partial {\partial k}
\ln k^{-p} J_p (k)  ;\nn
\eeq
the additional factor $k^{-p}$ has been introduced to avoid
contributions coming from the origin. Since no additional pole is enclosed,
the integral is unchanged.

The behaviour of $\zeta^{(n)} (s)$ as $n\to \infty$ controls the convergence of the
sum over $n$; different orders in $n$ can be
studied by shifting the contour to the imaginary axis and by using the
uniform asymptotic expansion of the resulting Bessel function $I_p (k)$.
To ensure that the resulting expression converges for some range of
$s$ when shifting the contour to the imaginary axis, we add a small positive constant
to the eigenvalues. For $s$ in the strip $1/2 < \Re s <1$, we have:
\beq 
\zeta ^{(n)} (s) = \frac{\sin (\pi s)} \pi
\int_\epsilon^\infty dk (k^2-\epsilon^2)^{-s}  \frac \partial {\partial k}
\ln k^{-p} I_p (k).\nn
\eeq

We introduce some additional notation dealing with the uniform asymptotic expansion of
the Bessel function. For $p\to\infty$ with $z=k/p$ fixed, we use results of
\cite{abra} to see that:
\begin{eqnarray}
&&I_p (zp) \sim \frac 1 {\sqrt{2\pi p}} \frac{e^{p\eta}}{(1+z^2)^{1/4}}
\left[ 1+\sum_{l=1}^\infty \frac{u_l (t)} { p^l} \right]\text{ where}\label{ex4}\\
&&t=1/\sqrt{1+z^2}\text{ and }\eta = \sqrt{1+z^2}+\ln [z/(1+\sqrt{1+z^2})].\nn
\end{eqnarray}
Let $u_0(t)=1$. We use the recursion relationship given in \cite{abra} to determine
the polynomials $u_l (t)$ which appear in equation (\ref{ex4}):
\beq 
u_{l+1} (t) = \frac 1 2 t^2 (1-t^2) u_l' (t) +\frac 1 8 \int_0^t d\tau
(1-5\tau^2) u_l (\tau).\nn
\eeq
We also need the coefficients $D_m (t)$ defined by the cumulant expansion:
\beq 
\ln \left[  1+\sum_{l=1}^\infty \frac{u_l (t)}{p^l} \right]
\sim \sum_{q=1}^\infty \frac{D_q (t)}{p^q} .\label{ex5}
\eeq
The eigenvalue multiplicities $d_n (m)$ are ${\mathcal O} (n^{m-2})$
as $n\to \infty$. Consequently, the leading behaviour of every term is on the order of
$p^{-2s-q+m-2}$; thus on the half plane $\Re s > (m-3)/2$, only the value $q=1$
contributes to the residues of the zeta-function. We have 
$
D_1 (t) = \frac 1 8 t -\frac 5 {24} t^3 
$. We use equation (\ref{ex4}) to decompose
\beq 
\zeta^ {(n)} (s) &=& A_{-1} ^{(n)} (s) + A_0 ^{(n)} (s) +
A_1 ^{(n)} (s) + R^{(n)} (s),\text{ where} \nn\\ 
A_{-1}^{(n)} (s) &=&  \frac{\sin \pi s}{\pi} \int_{\epsilon / p} ^\infty
dz [(zp)^2 -\epsilon^2] ^{-s} \frac {\partial}{\partial z} \ln
\left( z^{-p} e^{p\eta} \right) ,\nn\\ 
A_{0 }^{(n)} (s) &=&  \frac{\sin \pi s}{\pi} \int_{\epsilon / p} ^\infty
dz [(zp)^2 -\epsilon^2] ^{-s} \frac {\partial}{\partial z} \ln
\left( 1+z^2 \right)^{-1/4} ,\nn\\ 
A_{1}^{(n)} (s) &=&  \frac{\sin \pi s}{\pi} \int_{\epsilon / p} ^\infty
dz [(zp)^2 -\epsilon^2] ^{-s} \frac {\partial}{\partial z}
\left(\frac{D_1 (t) } p \right).\nn
\eeq
The remainder $R^{(n)}(s)$ is such that $ \sum_{n=0}^\infty d_n (m) R^{(n)} (s)$
is analytic on the half plane $\Re s > (m-3)/2$.

Let ${_2F_1}$ be the hypergeometric function. We have
\begin{eqnarray*}
&&{_2F_1} (a,b;c;z) = \frac{\Gamma (c) }{\Gamma (b) \Gamma (c-b)}
\int_0^1 dt t^{b-1} (1-t)^{c-b-1} (1-tz)^{-a},\text{ and}\\
&& \int_{\epsilon /p } ^\infty dz \,\, [(zp)^2 -\epsilon ^2] ^{-s} \frac
{\partial}{\partial z} t^l=
-\frac l 2 \frac {\Gamma (s+\frac l 2) \Gamma (1-s)}{\Gamma (1+\frac l 2) } p^l
[\epsilon^2 + p^2] ^{-s-l/2} .\nn
\end{eqnarray*}
We use the first identity to study $A_{-1}^{(n)}(s)$ and $A_0^{(n)}(s)$; we
use the second identity to study $A_1^{(n)}(s)$. This shows that
\beq
A_{-1}^{(n)} (s) &=&  \frac{\epsilon^ {-2s +1} } {2\Gamma( \frac{1}{2}) }
\frac{\Gamma (s-\frac 1 2)} {\Gamma (s)} {_2F_1}
(-{\frac{1}{2}},s-{\frac{1}{2}};
{\frac{1}{2}};-({\frac p \epsilon}
)^2 ) \nn  -\frac p 2
\epsilon ^{-2s} \nn
\\ A_0 ^{(n)} (s) &=&  -\frac 1 4 (p^2 + \epsilon ^2) ^{-s} , \nn \\
A_1 ^{(n)} (s) &=&  \frac 1 8 \frac 1 {\Gamma (s)} \left[
             -\frac{\Gamma (s+\frac{1}{2})}{\Gamma( \frac{1}{2}) }
            (p^2 +\epsilon ^2)^{-s-\frac{1}{2}}  \right]
                \nn\\
   & & -\frac 5 {24} \frac 1 {\Gamma (s)} \left[
-2\frac{\Gamma (s+\frac 3 2)} {\Gamma( \frac{1}{2}) }
 p^2 (p^2 +\epsilon ^2 )^{-s-\frac 3 2} \right].\nn
\eeq
We take the limit as $\epsilon \to 0$; the resulting zeta-function
which appears is connected to the spectrum on the sphere.
We define the
base zeta-function $\zeta_{S^d}$ and the Barnes zeta-function \cite{barnes} $\zeta_{{\mathcal B}}$
\begin{eqnarray*}
&& \zeta _{S^d} (s) = 4 \sum_{n=0}^\infty d_n (m) p^{-2s}\text{ and }
\zeta_{{\mathcal B}} (s,a) = \sum_{n=0}^\infty d_n (m) (n+a)^{-s}.\end{eqnarray*}
We then have the relation
$\zeta_{S^d} (s) = 2d_s \zeta_{{\mathcal B}} \left( 2s, \frac m 2 -1 \right)$. 
For $i=-1$, $i=0$, and $i=1$, let $A_i (s) = 4\sum_{n=0}^\infty d_n (m) A_i ^{(n)} (s)$.
We take the limit as $\epsilon \to 0$ to see that
\beq
&&A_{-1} (s)=   \frac 1 {4\Gamma( \frac{1}{2})} \frac {\Gamma (s-\frac 1 2) }{\Gamma
(s+1)}
\zeta_{S^d} (s-\frac 1 2) , \label{ex11} \\
&&A_0 (s)=   -\frac 1 4 \zeta_{S^d} (s) ,\text{ and} \label{ex12} \\
&&A_1 (s)= -\frac 1 {\Gamma (s) } \zeta_{S^d} (s+\frac 1 2) \left[
   \frac 1 {8\Gamma( \frac{1}{2})} \Gamma (s+\frac 1 2) -
  \frac 5 {12 \Gamma( \frac{1}{2}) } \Gamma (s+\frac3 2) \right]. \label{ex13}
\eeq

We use the Mellin-Barnes integral representation of the 
hypergeometric functions \cite{grad} to
calculate $A_{-1} (s)$: 
\beq 
{_2F_1} (a,b;c;z) = \frac{\Gamma (c)}{\Gamma (a) \Gamma (b)} \frac 1 {2\pi i}
\int_{{\mathcal C}} dt \,\, \frac{\Gamma (a+t) \Gamma (b+t) \Gamma (-t) }
{\Gamma (c+t)} (-z)^t.\nn
\eeq
We choose the contour of integration so that the poles of $\Gamma (a+t) \Gamma (b+t) /
\Gamma (c+t)$ lie to the left of the contour and so that the poles of $\Gamma (-t)$ lie to the
right of the contour. We stress that before interchanging the sum and the integral, we must
shift the contour ${\mathcal C}$ over the pole at $t=1/2$ to the left; this cancels
the term  $-\frac p 2 \epsilon ^{-2s}$ appearing in the expression for $A_{-1}$ appearing above.

This reduces the analysis of the
zeta function on the ball to analysis of a zeta function on the boundary. We compute the
residues of
$\zeta (s)$ from the residues of $\zeta _{{\mathcal B}}(s,a)$. Let $d:=m-1$. To compute
these residues, we first express $\zeta_{{\mathcal B}} (s,a)$ as a contour integral. Let
${\mathcal C}$ be the Hankel contour. 
\beq
\zeta_{{\mathcal B}} (s,a)
&=& \sum_{n=0}^\infty \left(
\begin{array}{c}
   d+n-1 \\
   n
\end{array}
\right) (n+a)^{-s} = \sum_{\vec m \in \nats_0^d} (a+m_1+...+m_d)^{-s}
\nn\\
&=& \frac{\Gamma (1-s) }{2\pi} \int_{{\mathcal C}} dt \,\, (-t)^{s-1}
\frac{e^{-at}} {(1-e^{-t})^d}.\nn
\eeq
The residues of $\zeta_{{\mathcal B}} (s,a)$
are intimately connected with the generalized Bernoulli polynomials
\cite{norlund},
\beq
\frac{e^{-at} } {(1-e^{-t} )^d} = (-1)^d
\sum_{n=0} ^\infty \frac{(-t)^{n-d} } {n!} B_n^{(d)} (a) .\label{ber}
\eeq
We use the residue theorem to see that
\beq 
\mbox{Res }_{s=z}\zeta_{{\mathcal B}} (s,a) = \frac{(-1)^{d+z} }{(z-1)! (d-z)!}
                    B_{d-z}^{(d)} (a) ,\label{barn}
\eeq
for $z=1,...,d$. The needed leading poles are
\beq 
\mbox{Res }_{s=d}\zeta_{{\mathcal B}} (s,a) &=&  \frac 1 {(d-1)!} ,\nn\\
 \mbox{Res }_{s=d-1} \zeta_{{\mathcal B}} (s,a) &=&
  \frac{d-2a}{2 (d-2)!} , \nn\\
 \mbox{Res }_{s=d-2} \zeta _{{\mathcal B}} (s,a) &=& 
 \frac {12 a^2 -d-12 a d +3d^2}
{24 (d-3)!} .\nn
\eeq
We may now determine the residues of
$\zeta (s)$. At $s=\frac m 2$, only $A_{-1} (s)$ contributes. We use
equation (\ref{ex11}) to see
\beq 
\mbox{Res }_{s=\frac m 2} \zeta (s) = d_s \frac 1 {4\sqrt{\pi }} \frac{\Gamma
\left(\frac{m-1} 2 \right) } {\Gamma \left( \frac m 2 +1 \right) }
\frac 1 {\Gamma (m-1)} .\nn
\eeq
We use the `doubling formula' $\frac{\Gamma (z)} {\Gamma (2z)} = \frac{\sqrt{2\pi} 2^{1/2-2z} }
{\Gamma (z+1/2)}\nn$ for the $\Gamma$ function, we use equation (\ref{ex1}), and we use
the observation $\Trace(1)=d_s$ to check that we have derived the correct formula
for $a_0$:
\beq 
a_0 = \frac{d_s}{2^m \Gamma \left( \frac m 2 +1 \right) }
= (4\pi )^{-m/2} \int_{B^m} \Trace (1).\nn
\eeq

Next, we study the pole at $s=\frac{m-1}2$. This time, $A_{-1}(s)$ and
$A_0 (s)$ contribute. We compute
\begin{eqnarray*}
&& \mbox{Res }_{s=\frac{m-1}2}A_{-1} (s)= \frac 1 {2^m \Gamma (\frac{m+1}2)
\Gamma (\frac{m-1}2)},
\nn\\
&& \mbox{Res }_{s=\frac{m-1}2}A_0 (s) = -\frac 1 {4\Gamma (m-1)}.\end{eqnarray*}
This implies that
\begin{eqnarray*}
&&a_{1}=\frac 1 {2^m \Gamma (\frac{m+1}2) }
-\frac{\Gamma( \frac{1}{2})} {2^m \Gamma (\frac m 2)}
=\frac 1 4 \left[\frac{\Gamma (\frac m 2) }{\Gamma( \frac{1}{2})
\Gamma (\frac{m+1}2) }
-1 \right] (4\pi)^{-\frac{m-1}2} \int_{S^{m-1}}  \Trace (1). \end{eqnarray*}
This determines $b_1(m)$ and proves the first assertion of Lemma \ref{THM4}.
Next we compute $a_2$ and thereby determine $c_3(m)$:
\begin{eqnarray*}
&& \mbox{Res }_{s=\frac{m-2}2} A_{-1} (s)= \frac 1 6 \,\,
\frac{4-m}{ 2^m} \frac 1 {\Gamma (\frac m 2)
\Gamma (\frac {m-2}2 )} ,\nn\\
&& \mbox{Res }_{s=\frac{m-2}2} A_0 (s) = -\frac 1 {8\Gamma (m-2)} ,\nn\\
&& \mbox{Res }_{s=\frac{m-2}2} A_1 (s) = -\frac 1 6 \,\,\frac{8-5m} {2^m} \frac 1
{\Gamma (\frac m 2) \Gamma (\frac {m-2} 2 )}.\end{eqnarray*}
We can now prove the second assertion of Lemma \ref{THM4} by determining $c_3(m)$. We
compute:
\begin{eqnarray*}
&&  a_2=\frac 2 3 \,\,
 \frac{(m-1)}{ 2^m \Gamma (\frac m 2) } -\frac 1 2 \,\,\frac{(m-1) \Gamma( \frac{1}{2})}
{ 2^m \Gamma (\frac{m+1}2)}\nn\\
&& \qquad= \frac 1 3 \left( 1-\frac{3\Gamma( \frac{1}{2})\Gamma (\frac m 2) }
{4\Gamma (\frac{m+1}2) }\right) (4\pi )^{-m/2} \int_{S^{m-1}} \Trace (L_{aa}).\nn
\end{eqnarray*}
The pattern of the universal constants involving a dimensionless constant and a
combination of $\Gamma$-functions is evident. This pattern holds for the
higher coefficients \cite{jon}.

We introduce the weighting (or smearing) function $F=1-r^2$ to
determine
$c_4(m)$; since $E$ and ${\mathcal R}$ vanish and since $F$ vanishes on the boundary,
only the term involving $F_{;m}=2$ survives. (Note: we checked these computations by
studying a more general function of the form
$F(r) = f_0 + f_1 r^2 +f_2 r^4$ but omit details in the interests of brevity).  We note
that the radial normalization constant is given by $C= 1/J_{p+1} (\lambda )$. We denote
the normalized Bessel function by
$$\bar J _k (\lambda r
) := J_k (\lambda r ) / J_{p+1} (\lambda ).$$
Instead of the
zeta function we consider now the smeared analogue:
\beq 
\zeta (F;s) = \sum_\lambda \int_{B^m} F(x) \varphi ^* (x) \varphi (x) \frac 1
{\lambda^{2s} }. \label{zetasmear}
\eeq
Since $F$ depends only on the normal variable, the integral in equation
(\ref{zetasmear}) over the sphere $S^{m-1}$ behaves as in the case $F=1$ so that
\begin{eqnarray*}
&& \zeta (F;s) = 4\sum_{n=0} ^\infty d_n (m) \int_{{\mathcal C}} \frac {dk}{2\pi i}
k^{-2s}\\ 
&& \qquad\qquad\cdot\int_0^1 dr F(r) r (\bar J^2_{p+1} (kr) + \bar J_p ^2 (kr)
)
\frac \partial {\partial k} \ln J_p (k) .\nn
\end{eqnarray*}
We set $F(r)=1-r^2$ and compute the radial integrals:
\beq
&& \int_0^1 r^3 \left[ \bar J ^2_p (\lambda r) + \bar J _{p+1} ^2 (\lambda r )
\right] =  \frac{2p^2 +3p +1} {3\lambda^2} +\frac 1 3 \text{ so}\nn\\
&& \zeta (r^2;s) = 4 \sum_{n=0}^\infty d_n (m)  \int_{{\mathcal C}} \frac {dk}{2\pi
i} k^{-2s} \left[  \frac{2p^2 +3p +1} {3k^2} +\frac 1 3 \right]
\frac \partial {\partial k} \ln J_p (k). \nn
\eeq
We use equation (\ref{ex2}) to evaluate this expression; the second term is given by
equation (\ref{ex2}) and simple substitutions suffice to evaluate the remaining parts.
The factor $1/k^2$ is absorbed by using $s+1$ instead of
$s$ in equations (\ref{ex11}), (\ref{ex12}), and (\ref{ex13}). The powers of $p$ lower
the argument of the base zeta-function by $1$, by $\frac 1 2$ and by $0$. It is now a
straightforward matter to compute:
\beq
A_{-1} (r^2;s) &=& 
\frac 1 {4\Gamma( \frac{1}{2})} \frac {\Gamma (s-\frac 1 2)}{\Gamma (s+1)} \zeta
_{S^d} (s-\frac 1 2) \left[\frac 1 3 +\frac 2 3 \frac{s-\frac 1 2}{s+1} \right] \nn\\
& & +\frac 1 {4\Gamma( \frac{1}{2})} \frac{\Gamma (s+\frac 1 2)}{\Gamma (s+2)}\left[
\zeta_{S^d} (s) +\frac 1 3 \zeta_{S^d} (s+\frac 1 2) \right] ,\nn\\
A_0 (r^2;s) &=& 
-\frac 1 4 \zeta _{S^d} (s) -\frac 1 4 \zeta _{S^d} (s+\frac 1 2)
+...\nn\\
A_1 (r^2;s) &=& 
-\frac 2 {3\Gamma (s+1)} \zeta_{S^d} (s+\frac 1 2)\left[
\frac 1 {8\Gamma( \frac{1}{2})} \Gamma (s+\frac3 2) -\frac 5 {12 \Gamma( \frac{1}{2}) }
\Gamma (s+\frac 5 2) \right]  \nn\\
& & -\frac 1 {3\Gamma (s)} \zeta_{S^d} (s+\frac 1 2)\left[
\frac 1 {8\Gamma( \frac{1}{2})} \Gamma (s+\frac 1 2) -\frac 5 {12 \Gamma( \frac{1}{2}) }
\Gamma (s+\frac 3 2) \right]+... \nn
\eeq
The coefficient $a_0$ gives the leading term in the expansion of the heat trace; this
 is correctly reproduced by the first term in
$A_{-1} (r^2;s)$. We also confirm the invariant form of the next coefficient
$a_{1}$. We determine the value of $c_4(m)$ by considering the pole at the value
$s=\frac{m-2}2$. Note that $F_{;m}=2$. We compute
\begin{eqnarray*}
&& \mbox{Res }_{s=\frac{m-2}2}\Gamma (s)A_{-1} (1-r^2;s)= -\frac 1
{12}
\frac m{m-2} (4\pi)^{-m/2} \int_{S^{m-1}}  \Trace (2)  \nn\\
&& 
\mbox{Res }_{s=\frac{m-2}2}\Gamma (s)A_0 (1-r^2;s)\nn\\
&& \qquad= \frac 1 4 \Gamma( \frac{1}{2}) \frac{m-1}{m-2}
\frac{\Gamma (\frac m 2)}{\Gamma ((m+1)/2)} 
(4\pi)^{-m/2} \int_{S^{m-1}}  \Trace (2)  \nn\\
&& 
\mbox{Res }_{s=\frac{m-2}2}\Gamma (s)A_1 (1-r^2;s)=-\frac 1 {12}\frac
{5m-6} {m-2} (4\pi)^{-m/2} \int_{S^{m-1}} \Trace (2) . \nn
\end{eqnarray*}
We sum the contributions to prove the third assertion of Lemma \ref{THM4} by evaluating
$c_4$. The final assertion of Lemma \ref{THM4} is now immediate.\ \qedbox

\section{Conclusion} In Section \ref{SEC2}, we used results of Grubb and Seeley to
study the setting when the structures are product near the boundary. In this setting,
$\hat\psi$ was self-adjoint and only the terms $\Trace(1)$ and $\Trace(\hat\psi)$ were
non-trivial. We used these results to determine the 
coefficients $b_1(m)$ and $c_0(m)$.

In Section \ref{SEC3}, we used functorial methods to determine the coefficients
$c_0(m)$, $c_1(m)$, and $c_2(m)$; the value we computed for $c_0(m)$ agreed with that
determined in Section \ref{SEC2}. We used variational methods to obtain a non-trivial
linear relationship between the coefficients $c_3(m)$ and $c_4(m)$. 

In
Section
\ref{SEC4}, we determined $b_1(m)$, $c_3(m)$ and $c_4(m)$ by studying the ball in
Euclidean space. The value of $b_1(m)$ computed in this fashion agreed with the value
which was determined in Section \ref{SEC2}. The values for the coefficients
$c_3(m)$ and $c_4(m)$ satisfied the linear relationship derived in Section \ref{SEC3}. 

In Section \ref{SEC3}, we chose $c=2$ to ensure that the $1$ parameter family
$\nabla(\varepsilon)$ of connections was unitary; the second author originally missed
this point and his error in choosing a different value of $c$ led to a seeming
contradiction, which has now been cleared up, between the methods of Section \ref{SEC3}
and the methods of Section \ref{SEC4}. We find it interesting that each of
the three methods we have discussed gives some, but not all, of the coefficients and
that none of the results of the three sections is a proper subset of the other.

\end{document}